\documentstyle[12pt,epsfig,amssymb]{article}
\begin{document}
\setlength{\baselineskip}{0.75cm}
\setlength{\parskip}{0.45cm}
\begin{titlepage}
\begin{flushright}
DO-TH 97/02 \linebreak
February 1997 
\end{flushright}
\vskip 1.2in
\begin{center}
{\Large\bf Charged-Current Leptoproduction of D-Mesons\\ 
  in the Variable Flavor Scheme}
\vskip 0.8in
{\large S.\ Kretzer and I.\ Schienbein}
\end{center}
\vskip .3in
\begin{center}
{\large Institut f\"{u}r Physik, Universit\"{a}t Dortmund \\
D-44221 Dortmund, Germany}
\vskip 0.3in
{\it{To appear in Physical Review D}}
\vskip 0.4in
\end{center}
{\large{\underline{Abstract}}}

\noindent
We present formulae for the momentum ($z$) distributions of D-mesons
produced in neutrino deep-inelastic scattering off strange partons. The 
expressions are derived within the variable flavor scheme of Aivazis
et al.\ (ACOT-scheme), which is extended from its fully inclusive
formulation to one-hadron inclusive leptoproduction.   
The dependence of the results on the assumed strange quark
mass $m_s$ is investigated and the $m_s \to 0$ limit is compared to
the corresponding $\overline{\rm{{MS}}}$ results. The importance of 
${\cal{O}}(\alpha_s)$ quark--initiated corrections is demonstrated for the
$m_s=0$ case.
\end{titlepage}
 
\noindent
The momentum ($z$) distributions of D-mesons from the fragmentation of 
charm quarks produced in neutrino deep-inelastic scattering (DIS) have 
been used recently to determine the strange quark distribution of the 
nucleon $s(x,Q^2)$ at leading order (LO) \cite{ref1} and next-to-leading 
order (NLO) \cite{ref2}. A proper QCD calculation of this quantity 
requires the convolution of a perturbative hard scattering charm production
cross section with a nonperturbative $c \rightarrow D$ fragmentation 
function $D_c(z)$ leading at ${\cal{O}}(\alpha_s)$ to the breaking of 
factorization in Bjorken-$x$ and $z$ as is well known for light quarks 
\cite{ref3}. So far experimental analyses have assumed a factorized cross 
section even at NLO \cite{ref2}. This shortcoming has been pointed out in 
\cite{ref4} and the hard scattering convolution kernels needed for a 
correct and complete NLO analysis have been calculated in the  
$\overline{\rm{{MS}}}$ scheme with three massless flavors ($u,d,s$) using 
dimensional regularization. In the experimental NLO analysis in \cite{ref2}
the variable flavor scheme (VFS) of Aivazis, Collins, Olness and Tung 
(ACOT) \cite{ref5} for heavy flavor leptoproduction has been utilized. 
In this formalism one considers, in addition to the quark scattering (QS) 
process, e.g.\  $W^+ s \rightarrow c$, the contribution from the gluon 
fusion (GF) process $W^+ g \rightarrow c \bar{s}$ with its full 
$m_s$--dependence. The collinear logarithm  which is already contained in 
the renormalized $s(x,Q^2)$ is subtracted off numerically. The 
quark--initiated contributions from the subprocess $W^+ s \rightarrow c g$ 
(together with virtual corrections) which were included in the complete 
NLO ($\overline{\rm{{MS}}}$) analysis in \cite{ref4} are usually neglected
in the ACOT--formalism. The ACOT--formalism has been formulated explicitly 
only for fully inclusive leptoproduction \cite{ref5}. It is the main 
purpose of this article to fill the gap and provide the expressions needed 
for a correct calculation of one-hadron (D-meson) inclusive leptoproduction
also in this formalism.

In the following we will stick closely to the ACOT formalism as formulated
in \cite{ref5} except that we are not working in the helicity basis 
but prefer the standard tensor basis implying the usual structure functions
$F_{i=1,2,3}$. We are not considering kinematical effects arising from an 
initial state quark mass in the $W^+ s \rightarrow c$ quark scattering 
contribution, i.e., $s(x,Q^2)$ represents massless initial state strange 
quarks. This latter choice must be consistently kept in the subtraction 
term \cite{ref5} to be identified below from the $m_s \to 0$
limit of the $W^+ g \rightarrow c \bar{s}$ gluon fusion contribution.
The fully massive partonic matrix elements have been calculated for the 
general boson-gluon-fusion process $B g \rightarrow {\bar{Q}}_1 Q_2$ 
in \cite{ref6} where $B=\gamma^*,\ W^{\pm},\ Z$. When they are 
convoluted with a nonperturbative gluon distribution $g(x,\mu^2)$ and a 
fragmentation function $D_{Q_2}(z)$, one obtains the GF part of the 
hadronic structure function $F_i(x,z,Q^2)$ describing the momentum ($z$) 
distribution of a hadron $H$ containing the heavy quark $Q_2$:
\begin{eqnarray}  \nonumber
F_{1,3}^{GF}(x,z,Q^2) &=& \int_{ax}^{1} \frac{dx'}{x'} 
              \int_{max[z,\zeta_{min}(x/x')]}^{\zeta_{max}(x/x')}
      \ \frac{d\zeta}{\zeta}\ g(x',\mu^2)\ f_{1,3}(\frac{x}{x'},\zeta,Q^2)
      \ D_{Q_2}(\frac{z}{\zeta})  \\ \nonumber \\
F_{2}^{GF}(x,z,Q^2) &=& \int_{ax}^{1} \frac{dx'}{x'} 
              \int_{max[z,\zeta_{min}(x/x')]}^{\zeta_{max}(x/x')}
      \ \frac{d\zeta}{\zeta}\ x'g(x',\mu^2)\ f_{2}(\frac{x}{x'},\zeta,Q^2)
      \ D_{Q_2}(\frac{z}{\zeta})
\end{eqnarray}
\noindent
with the fractional momentum variables  $z=p_H\cdot p_N / q\cdot p_N$
and $\zeta =p_{Q_2}\cdot p_N / q\cdot p_N$, $p_N$ and $q$ being the 
momentum of the nucleon and the the virtual boson, respectively. 
The structure functions $F_i(x,z,Q^2)$ generalize the usual fully inclusive
structure functions $F_i(x,Q^2)$, if one considers one-hadron (H) inclusive
leptoproduction. The partonic structure functions $f_i(x',\zeta,Q^2)$ are 
given by
\begin{equation}
f_{i=1,2,3}\left(x',\zeta,Q^2\right)=
\frac{\alpha_s(\mu^2)}{\pi}\ \left[\ \frac{A_i}
{(1-\zeta)^2}+\frac{B_i}{\zeta^2}+\frac{C_i}{1-\zeta}
+\frac{D_i}{\zeta}+E_i\ \right] 
\end{equation}
\noindent
with
\begin{eqnarray*}
A_1\left(x',Q^2\right)&=&q_{+}\ \frac{{x'}^2}{4}\ \frac{m_1^2}{Q^2}
\ \left(\ 1+
\frac{\Delta m^2}{Q^2}-\frac{q_{-}}{q_{+}}\frac{2 m_1 m_2}{Q^2}\right)\\
C_1\left(x',Q^2\right)&=&\frac{q_{+}}{4}\ \Bigg[\ \frac{1}{2}-x'(1-x')
-\frac{\Delta m^2\,x'}
{Q^2}(1-2x')+\left(\frac{\Delta m^2\,x'}{Q^2}\ \right)^2  \\
&+& \frac{q_{-}}{q_{+}}\ \frac{m_1 m_2}{Q^2}
\ 2x'\ (1-x'-x'\ \frac{m_1^2+m_2^2}{Q^2})\ \Bigg] \\
E_1\left(x',Q^2\right)&=&\frac{q_{+}}{4}\ (\ -1+2x'-2{x'}^2\ )\\
A_2\left(x',Q^2\right)&=&q_{+}\ x'\ \left[{x'}^2\ \frac{m_1^2}{Q^2}
\left(\frac{1}{2}
\left(\frac{\Delta m^2}{Q^2}\right)^2+\frac{\Delta m^2\,-m_1^2}{Q^2}
+\frac{1}{2}\right)\right]\\
C_2\left(x',Q^2\right) &=& q_{+}\ \frac{x'}{4}\ \Bigg[\ 1-2x'(1-x')
+\frac{m_1^2}{Q^2}
\left(1+8x'-18{x'}^2\right) \\
&+& \frac{m_2^2}{Q^2}\left(1-4x'+6{x'}^2\right) 
- \frac{m_1^4+m_2^4}{Q^4}\ 2x'(1-3x')+\frac{m_1^2 m_2^2}{Q^4}\ 4x'(1-5x')\\
&+&\frac{\Delta m^4\,\Delta m^2}{Q^6}2{x'}^2 
- \frac{q_{-}}{q_{+}}\ \frac{2 m_1 m_2}{Q^2}\ \Bigg] \\
E_2\left(x',Q^2\right)&=&q_{+}\ x'\ \left[-\frac{1}{2}+3x'(1-x')\right]\\
A_3\left(x',Q^2\right)&=&R_q\ m_1^2\ {x'}^2\ \frac{\Delta m^2 +Q^2}{Q^4}\\
C_3\left(x',Q^2\right)&=&R_q\left[\ \frac{1}{2}-x'(1-x')
-\frac{\Delta m^2}{Q^2}
\ x'(1-2x')+\frac{\Delta m^4}{Q^4}\ {x'}^2\right]\\
E_3\left(x',Q^2\right)&=&0 \\
B_{i={1,2 \atop 3}}\left(x',Q^2\right)&=&\pm A_i\left(x',Q^2\right)
[m_1\leftrightarrow m_2]\\
D_{i={1,2 \atop 3}}\left(x',Q^2\right)&=&\pm C_i\left(x',Q^2\right)
[m_1\leftrightarrow m_2]
\end{eqnarray*}
\noindent
where $\Delta m^n \equiv m_2^n-m_1^n\ ,\ m_{1,2}$ being the mass of
the heavy quark $Q_{1,2}$. The kinematical boundaries of phase space 
in the convolutions in eq.\ (1) are
\begin{equation}
a x = \left[1+\frac{(m_1+m_2)^2}{Q^2}\right]\ x\ \ \ ,\ \  
\ \zeta_{min,max}(x')=\frac{1}{2} \left[ 1+
\frac{\Delta m^2}{Q^2} \frac{x'}{1-x'} \pm v {\bar{v}} \right]
\end{equation}
\noindent
with $ \displaystyle \quad
v^2=1-\frac{(m_1+m_2)^2}{Q^2} \frac{x'}{1-x'} \ \ ,    
\ \ {\bar{v}}^2=1-\frac{(m_1-m_2)^2}{Q^2} \frac{x'}{1-x'} \quad $.
The vector ($V$) and axialvector ($A$) couplings of the 
$\gamma_{\mu}(V-A \gamma_5)$ quark current enter via 
$q_{\pm}=V^2\pm A^2$, $R_q=V A$ \cite{ref6b}.
If the partonic structure functions in eq.\ (2) are integrated over
$\zeta$ the well known inclusive structure functions \cite{ref7,ref7b} 
for heavy flavor production are recovered:
\begin{equation}
\int_{\zeta_{min}(x')}^{\zeta_{max}(x')} d\zeta
\ f_{i={1,2 \atop 3}}(x',\zeta,Q^2) = \pm f_i(x',Q^2)
\ \ \ ,
\end{equation}
where the   $f_i(x',Q^2)$ can be found in \cite{ref7}.

In the following we will consider the special case of charged current
charm production, i.e., $m_1=m_s$, $m_2=m_c$ ($q_{\pm}=2,0$; $R_q=1$
assuming a vanishing Cabibbo angle). Of course, all formulae below can be 
trivially adjusted to the general case of eqs.\ (1,2).
The $m_s \to 0$ limit of the partonic structure functions in 
eq.\ (2) is obtained by keeping terms up to ${\cal{O}}(m_s^2)$  in the 
$A_i$,
$C_i$ and in $\zeta_{max}$ due to the singularity of the phase space
integration stemming from  $\zeta \to 1$. One obtains
\begin{eqnarray} \nonumber
\lim_{m_s \to 0}\ 
\frac{\pi}{\alpha_s}\ f_{i}(x',\zeta,Q^2) &=&
c_i\ H_i^g(\frac{x'}{\lambda},\zeta,m_s^2,\lambda) \\
&=& c_i\ \delta (1-\zeta )\ P_{qg}^{(0)}(\frac{x'}{\lambda}) 
\ \ln \frac{Q^2+m_c^2}{m_s^2} + {\cal{O}}(m_s^0)
\end{eqnarray}
\noindent
where $\displaystyle P_{qg}^{(0)}(x') = \frac{1}{2} [{x'}^2+(1-{x'})^2]$, 
$\lambda=Q^2/(Q^2+m_c^2)$, $c_1=1/2$, $c_2=x'/\lambda$ ,$c_3=1$
and the $H_i^g$ are the same as the dimensionally regularized 
$\overline{\rm{{MS}}}$ ($m_s=0$) gluonic 
coefficient functions obtained in \cite{ref4}. The $c_i$ arise from 
different normalizations of the $f_i$ and the $H_i^g$ and are such that
the infrared--safe subtracted [see below eq.(7)] convolutions in eq.\ (1)
converge towards the corresponding ones in \cite{ref4} as
$m_s \to 0$ if one realizes
that  $x'/\lambda=\xi'$, $x/\lambda=\xi\equiv x(1+m_c^2/Q^2)$.
Taking also the limit $m_c \to 0$ in eq.\ (5) gives --besides the 
collinear logarithm already present in eq.\ (5)-- finite expressions
which agree \cite{ref7c} with the massless results of \cite{ref3}. 

In the ACOT formalism the GF convolutions in eq.\ (1) coexist with the
Born level quark scattering contributions 
$F_i^{QS}(x,z,Q^2) = k_i\ s(\xi,\mu^2)\ D_c(z)$, 
$k_{i=1,2,3}=1,\ 2\xi,\ 2$. The overlap between the QS and the GF 
contributions is removed by introducing a subtraction term (SUB) 
\cite{ref5} which is obtained from the massless limit in eq.\ (5) 
\begin{equation}
F_i^{SUB}=k_i\  \frac{\alpha_s(\mu^2)}{2\pi}\ \ln\frac{\mu^2}{m_s^2}\
\left[
\int_{\xi}^1 \frac{dx'}{x'}\ g(x',\mu^2)
\ P_{qg}^{(0)}\left(\frac{\xi}{x'}\right)\right]\ D_c(z)\ \ \ .
\end{equation}
The complete ${\cal{O}}(\alpha_s)$ structure functions for the $z$
distribution of charmed hadrons (i.e., dominantly D-mesons) produced
in charged current DIS are then given in the ACOT formalism \cite{ref5} by
\begin{equation}
F_i^{ACOT}=F_i^{QS}-F_i^{SUB}+F_i^{GF}\ \ \ .
\end{equation}

In Fig.\ 1 we show the structure function $F_2^{ACOT}$ at experimentally 
relevant \cite{ref1} values of $x$ and $Q^2$  
for several finite choices of $m_s$ 
together with the asymptotic $m_s \to 0$ limit. For $D_c$  we use
a Peterson fragmentation function \cite{ref8} 
\begin{equation}
D_c(z) = N \left\{ z \left[ 1-z^{-1}-\varepsilon_c/(1-z)
\right]^2\right\}^{-1}
\end{equation}
with $\varepsilon_c=0.06$ \cite{ref9,ref10} normalized to 
$\int_0^1 dz D_c(z) = 1$ 
and we employ the GRV94(HO) parton distributions \cite{ref11} with
$m_c=1.5\ {\rm{GeV}}$. Our choice of 
the factorization scale is $\mu^2=Q^2+m_c^2$ which ensures that there is
no large $\ln (Q^2+m_c^2)/\mu^2$ present in the difference GF--SUB.    
As can be seen from Fig.\ 1 the effects of a finite
strange mass are small and converge rapidly towards the massless
$\overline{\rm{{MS}}}$ limit provided $m_s \lesssim 200\ {\rm{MeV}}$ as
is usually assumed \cite{ref2}.

In Fig.\ 2 we show the effects of adding the quark--initiated 
${\cal{O}}(\alpha_s)$ correction 
from the process  $W^+ s \rightarrow c g$ (together with virtual 
corrections) to the asymptotic ($m_s\to 0$) $F_2^{ACOT}$,
employing the CTEQ4($\overline{\rm{{MS}}}$)
densities \cite{ref12} with $m_c=1.6\ {\rm{GeV}}$.
The  ${\cal{O}}(\alpha_s)$ quark contribution is usually neglected in the 
ACOT formalism since 
it is assumed to be effectively suppressed by one order of 
$\alpha_s$ with respect to the gluon fusion contribution due to
$s(x,\mu^2)/g(x,\mu^2) = {\cal{O}}(\alpha_s)$. To check this assumption
for the quantity $F_2(x,z,Q^2)$ we show, besides the full result, the 
contributions from the different subprocesses 
(using again $\mu^2=Q^2+m_c^2$).
The $W^+ g \rightarrow c {\bar{s}}$ contribution corresponds to 
GF--SUB in eq.\ (7).
The quark--initiated ${\cal{O}}(\alpha_s)$ contribution has been calculated
in the $\overline{\rm{{MS}}}$ scheme according to \cite{ref4} 
which is consistent with the 
asymptotic gluon--initiated correction in the ACOT scheme due to eq.\ (5).
It can be seen that the quark--initiated correction is comparable in size
to the gluon--initiated correction around the maximum of $F_2$. Since 
most of the experimentally measured D-mesons originate from this region
the ${\cal{O}}(\alpha_s)$ quark contributions should not be neglected in a 
complete NLO calculation. It would be worthwhile to calculate these 
diagrams also in the ACOT scheme to study finite $m_s$ effects as
has been done in this article for the ${\cal{O}}(\alpha_s)$ gluon 
contributions.

To summarize we have given formulae which extend the ACOT scheme 
\cite{ref5} for the
leptoproduction of heavy quarks from its fully inclusive formulation 
to one-hadron inclusive leptoproduction. We have applied 
this formulation to 
D-meson production in charged current DIS and studied finite $m_s$
corrections to the asymptotic $m_s \to 0$ limit. The corrections 
turned out to be small for reasonable choices of 
$m_s \lesssim 200\ {\rm{MeV}}$ and we have shown that the  
$m_s \to 0$ limit  
reproduces the dimensionally regularized $\overline{\rm{{MS}}}$
($m_s=0$) gluonic coefficient functions \cite{ref4}.   
Furthermore we have investigated the
quark--initiated ${\cal{O}}(\alpha_s)$ corrections for $m_s=0$ using
the relevant $\overline{\rm{{MS}}}$ fermionic coefficient functions
\cite{ref4}. The latter corrections turned out to be numerically important 
at experimentally relevant values of $x$ and $Q^2$ \cite{ref1} and should
be included in a complete NLO calculation of charged current 
leptoproduction of D-mesons.
\section*{Acknowledgements}
We thank E.\ Reya and M.\ Gl\"{u}ck for advice and useful discussions
and E.\ Reya for carefully reading the manuscript.
This work has been supported in part by the
'Bundesministerium f\"{u}r Bildung, Wissenschaft, Forschung und
Technologie', Bonn.    

\newpage
%
%
\newpage
\section*{Figure Captions}
\begin{description}
\item[Fig.\ 1] The structure function $F_2^{ACOT}(x,z,Q^2)$ as defined
in eq.\ (7) using the GRV94(HO) parton densities \cite{ref11} with 
$m_c=1.5\ {\rm{GeV}}$ and a Peterson fragmentation function \cite{ref8}
with $\varepsilon_c = 0.06$. Several finite choices for $m_s$ are shown as
well as the asymptotic $m_s \to 0$ limit.    
\item[Fig.\ 2] The structure function $F_2(x,z,Q^2)$ for charged current
leptoproduction of D-mesons at ${\cal{O}}(\alpha_s)$ for $m_s=0$ using
the CTEQ4($\overline{\rm{{MS}}}$) parton distributions \cite{ref12} and
a Peterson fragmentation function \cite{ref8} with $\varepsilon_c = 0.06$.
The full
${\cal{O}}(\alpha_s)$ result is shown as well as the individual 
contributions from the distinct quark-- and gluon--initiated processes.
\end{description}
\newpage
\pagestyle{empty}

\begin{figure}
\vspace*{-1cm}

\hspace*{-1.5cm}
\epsfig{figure=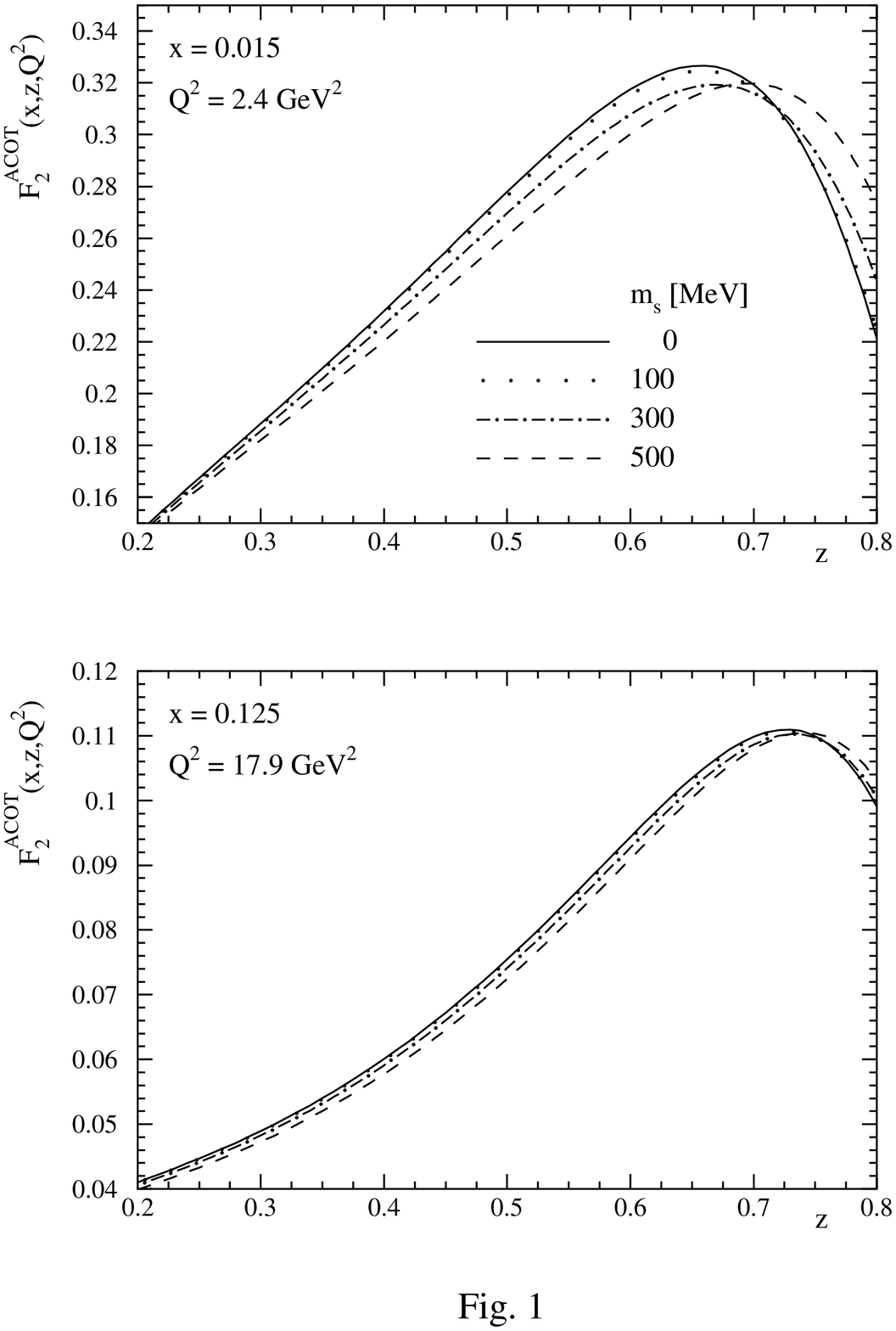,width=20cm}
\end{figure}
\newpage
\begin{figure}
\vspace*{-1.5cm}

\hspace*{-2.5cm}
\epsfig{figure=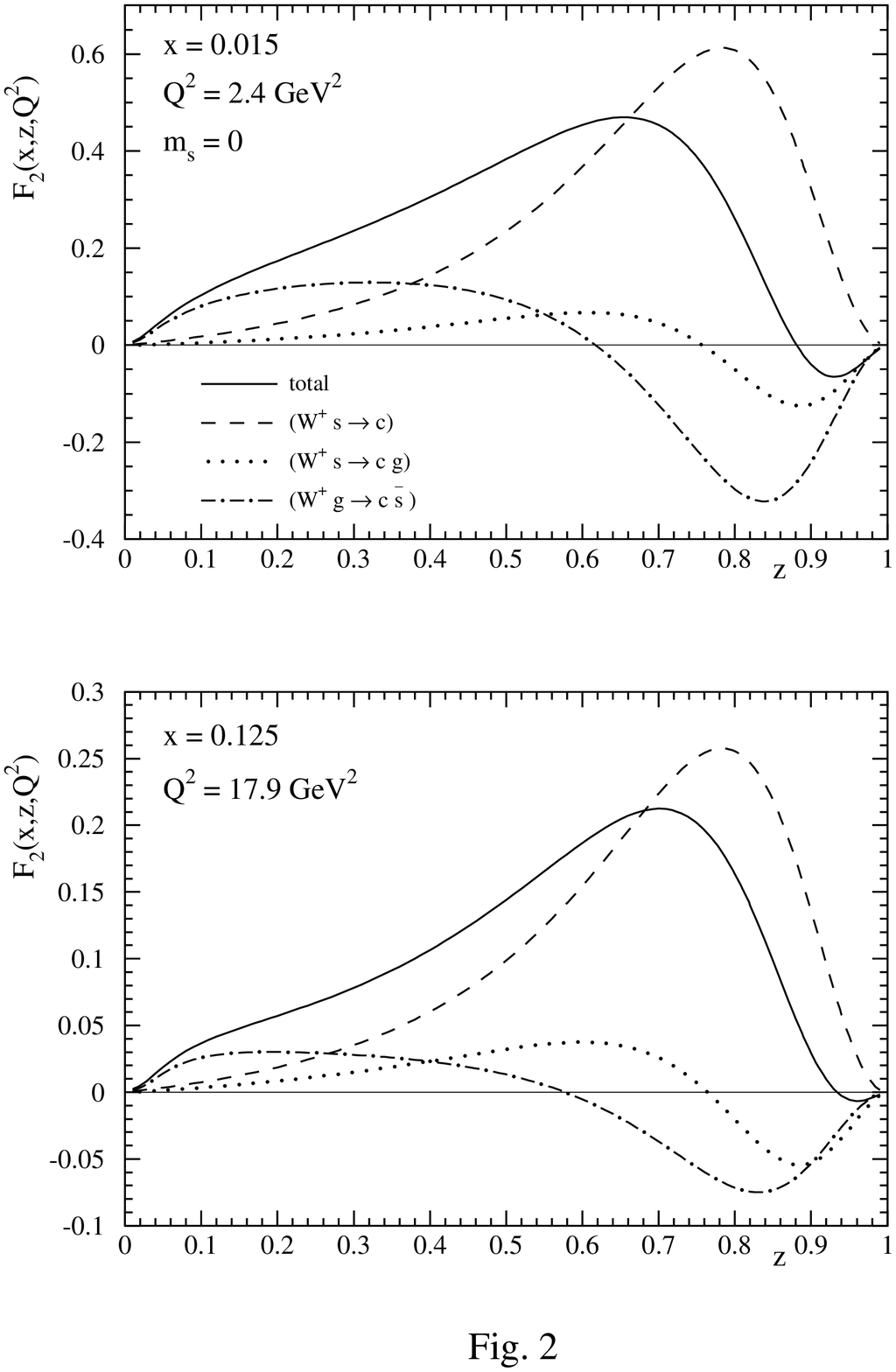,width=20cm}
\end{figure}

\begin{thebibliography}{99}
\bibitem{ref7c}
This is slightly different for the fermionic coefficient functions in 
\cite{ref4} which have a finite $m_c \to 0$ limit that differs from
the massless results in \cite{ref3} by terms $\sim \delta (1-x')$.
The difference is due to $\zeta_{min} \to 1$ as $\xi ' \to 1$ 
such that terms
$\sim \delta (1-\xi ')$ are automatically $\sim \delta (1-\zeta)$
\cite{ref4} if $m_c \neq 0$.
\end{thebibliography}
\end{document}